\documentclass{aa}  

\usepackage{graphicx}
\usepackage{txfonts}
\usepackage{siunitx,subfig,xspace,natbib}
\usepackage{hyperref}
\usepackage{amsmath}%
\usepackage{amsfonts}%
\usepackage{amssymb}%
\usepackage{mathrsfs}
\usepackage{arydshln}
\usepackage{cprotect}
\usepackage{multirow}

\DeclareMathOperator\erfc{erfc}

\newcommand{\mrm}[1]{\mathrm{#1}}

\begin{document}

   \title{Background modelling for $\gamma$-ray spectroscopy with INTEGRAL/SPI}


\author{
  Thomas Siegert   		     \inst{\ref{inst:mpe},\ref{inst:xcu}}\thanks{E-mail: tsiegert@mpe.mpg.de} \and
	Roland Diehl             \inst{\ref{inst:mpe},\ref{inst:xcu}} \and	
  Christoph Weinberger     \inst{\ref{inst:mpe}} \and
	Moritz M. M. Pleintinger \inst{\ref{inst:mpe}} \and
	Jochen Greiner           \inst{\ref{inst:mpe},\ref{inst:xcu}} \and
	Xiaoling Zhang           \inst{\ref{inst:mpe}}
}
\institute{
  Max-Planck-Institut f\"ur extraterrestrische Physik, Gie\ss enbachstra\ss e, D-85741 Garching, Germany
  \label{inst:mpe}
  \and
  Excellence Cluster Universe, Boltzmannstra\ss e 2, D-85748, Garching, Germany
  \label{inst:xcu}
  }

   \date{Received December 19, 2018; accepted February 25, 2019}

 
  \abstract
   {The coded-mask spectrometer-telescope SPI on board the INTErnational Gamma-Ray Astrophysics Laboratory (INTEGRAL) records photons in the energy range between 20 and 8000~keV. A robust and versatile method to model the dominating instrumental background (BG) radiation is difficult to establish for such a telescope in the rapidly changing space environment.}
   {From long-term monitoring of SPI's Germanium detectors, we built up a spectral parameter data base \citep{Diehl2018_BGRDB}, which characterises the instrument response as well as the BG behaviour. We aim to build a self-consistent and broadly applicable BG model for typical science cases of INTEGRAL/SPI, based on this data base.}
   {The general analysis method for SPI relies on distinguishing between illumination patterns on the 19-element Germanium detector array from BG and sky in a maximum likelihood framework. We illustrate how the complete set of measurements, even including the exposures of the sources of interest, can be used to define a BG model. The observation strategy of INTEGRAL makes it possible to determine individual BG components, originating from continuum and $\gamma$-ray line emission. We apply our method to different science cases, including point-like, diffuse, continuum, and line emission, and evaluate the adequacy in each case.}
   {From likelihood values and the number of fitted parameters, we determine how strong the impact of the unknown BG variability is. We find that the number of fitted parameters, i.e. how often the BG has to be re-normalised, depends on the emission type (diffuse with many observations over a large sky region, or point-like with concentrated exposure around one source), and the spectral energy range and bandwidth. A unique time scale, valid for all analysis issues, is not applicable for INTEGRAL/SPI, but must and can be inferred from the chosen data set.}
   {We conclude that our BG modelling method is usable in a large variety of INTEGRAL/SPI science cases, and provides nearly systematics-free and robust results.}

   \keywords{Gamma rays: general; Methods: data analysis; Techniques: spectroscopic}

   \maketitle
%

\section{Introduction}

Since 2002, the spectrometer SPI \citep{Vedrenne2003_SPI} on board the INTErnational Gamma-Ray Astrophysics Laboratory \citep[INTEGRAL satellite][]{Winkler2003_INTEGRAL} has been observing astrophysical high-energy phenomena in the hard X-ray and soft $\gamma$-ray range between 20~keV and 8~MeV. In this energy range, measured $\gamma$-ray spectra are dominated by instrumental background (BG) photons. These originate mainly from nuclear de-excitation reactions and continuum processes, such as bremsstrahlung, of the instrument and satellite material, being exposed to cosmic-rays (CRs). There is no stand-alone BG model available to deal with SPI observations in an extensive, consistent, and elaborate way. In this paper, we show how to construct a self-consistent BG model for SPI data analysis from a data base of spectral parameters over the INTEGRAL mission years.

The spectra in each of the 19 high-purity Ge detectors of SPI can be characterised by a large number of instrumental $\gamma$-ray lines on top of a broken power-law shaped continuum. The main features of the spectra among detectors, energy, and time stay constant and change gradually according to solar activity and detector degradation. \citet{Diehl2018_BGRDB} illustrated how the SPI spectral BackGround Response Data Base (BGRDB) is created, maintained, and checked for consistency over the entire mission time. In general, the BGRDB contains fitted spectral parameters per detector and time. The time integration is chosen as either one INTEGRAL orbit of three\footnote{In early 2015, the INTEGRAL spacecraft performed several orbit adjustment manoeuvre, to safely bring the satellite back to Earth in 2029. For this reason, the INTEGRAL orbit is now merely 2.7~days long, and will decrease with ongoing mission time.} days, or the time between two detector restoration\footnote{About twice a year, the lattice structure of the SPI Germanium detectors is repaired from cosmic-ray radiation by heating up the camera for several days to $\approx 100~\mrm{^{\circ}C}$. This is called an annealing.} periods, which is typically half a year. These different time scales are a key issue for the understanding and the construction of a $\gamma$-ray telescope BG model - to record sufficient BG data for reliable fits to the spectral, and to be able to trace gradual changes in the spectral response.

This paper is structured as follows: In Sec.~\ref{sec:data_analysis}, we describe, how SPI data analysis is generally treated (Sec.~\ref{sec:general_method}), and describe the parts of the SPI BGRDB which are required for BG modelling in more detail. Based on this, we explain how we construct a self-consistent BG model in Sec.~\ref{sec:bg_model_construction}, with full algorithm detail (Sec.~\ref{sec:general_approach}) and a discussion about the underlying foundation (Sec.~\ref{sec:why_does_this_work}). In Sec.~\ref{sec:fit_adequacy}, we show how to evaluate the BG model fit adequacy for different test cases for point-like and diffuse emission (Sec.~\ref{sec:data_sets}), which are then further discussed considering their temporal BG behaviour (Sec.~\ref{sec:time_scale}) in Sec.~\ref{sec:test_cases}.

\begin{figure*}[!ht]
	\centering
	
		\subfloat[Typical $5\times5$ rectangular dithering strategy with SPI, marked with dots and sequenced by number above the insets. Each dot represents one pointing with a field of view of $16^{\circ} \times 16^{\circ}$ and is $2.1^{\circ}$ away from the next/neighbouring pointing. The celestial source is marked with a blue star symbol at the position $(l/b) = (35.8^{\circ} / -3.9^{\circ})$. In each inset panel, ranging from 0 to 2.75, the relative detector pattern of how the source would be seen by SPI is shown. The dashed lines in each sub-panel indicate how different the expected sky pattern appears and changes from pointing to pointing with respect to a flat $1:1:...:1$-ratio pattern. \label{fig:sky_pattern}]{\includegraphics[width=1.4\columnwidth,trim=1.0cm 1.1cm 2.4cm 2.1cm,clip=true]{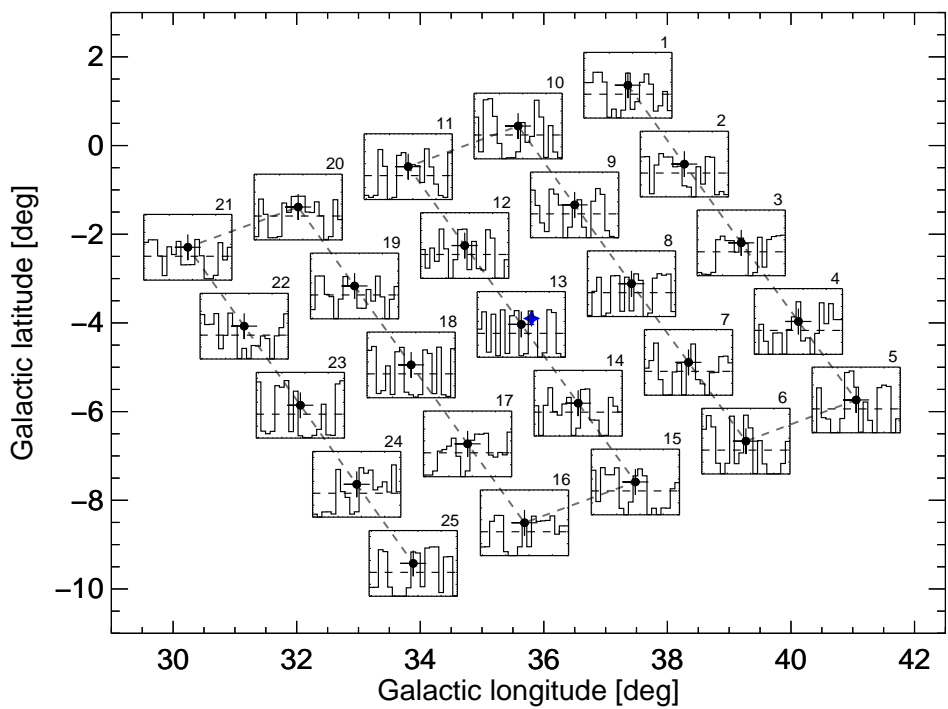}}\\
	
	  \subfloat[Zoom of panel 13 in Fig.~\ref{fig:sky_pattern}. The relative detector pattern is shown; the dashed line at 1.0 marks a flat detector pattern. \label{fig:pat_pan13}]{\includegraphics[width=0.813\columnwidth,trim=2.5cm 1.9cm 2.4cm 2.4cm,clip=true]{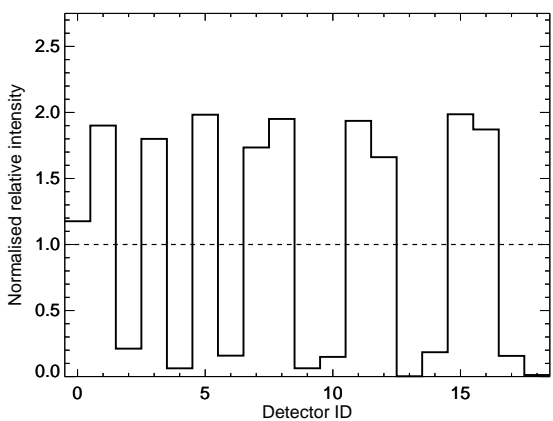}}~~~~
	  \subfloat[Shadowgram equivalent to detector pattern of panel 13. \label{fig:dets_pan13}]{\includegraphics[width=0.687\columnwidth,trim=2.8cm 0.8cm 2.8cm 0.8cm,clip=true]{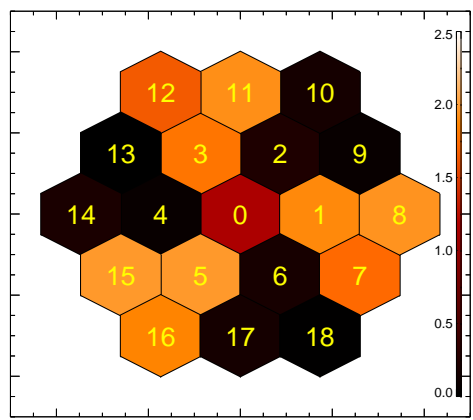}}
	  \caption{Detector pattern and shadowgram of a celestial source near the optical axis of SPI. From \citet{Siegert2017_PhD}.}
\end{figure*}

\section{SPI data analysis}\label{sec:data_analysis}

\subsection{General method - fitting time series}\label{sec:general_method}

SPI data analysis and source flux extraction is performed for individual energy bins. Each energy bin\footnote{The smallest public available bin size in SPI data analysis is 0.5~keV. The chosen bin width for the actual data set, however, is arbitrary. To increase the statistics and to reduce the time for the analysis at the same time (number of bins), for example, the bin width can be increased. Note, that spectral information is lost when large energy bins are used.} is treated separately, even though the bins are connected, e.g. by instrumental resolution. This dependence is taken into account when the BG model is built.

The counting of photons into one energy bin over time obeys the Poisson statistics. From this, the likelihood, $\mathcal{L}(\boldsymbol{\theta}|D)$, can be calculated, given the data set, $D$, and the model parameters $\boldsymbol{\theta}$. The data in SPI per energy bin is structured as a vector of magnitude $\mrm{"number~of~detectors"}\times\mrm{"number~of~pointings"} = \left\|d\right\| \times \left\|p\right\|$, where $\left\|d\right\|$ would be equal to 19 if all individual detectors\footnote{SPI also records events which occur on multiple detectors within a short time, classified as double, triple, and high-order detector hits, which are individually saved in SPI data sets.} of SPI are used, and $\left\|p\right\|$ equals the number of observations, $N_{obs}$, in a specific data set.

A pointing, $p$, is an observation unit for a specific amount of time, $T_p$, typically 0.5 to 1.0~h, and a specific region in the sky, marked by the galactic longitude and latitude $(l_p/b_p)$. The satellite is staring during this time at this position, and is recording data. Depending on the type of observation (diffuse/point-like; persistent/transient), $N_{obs}$ pointings are observed over larger regions in the sky, or only around the target source. Sub-pointing time intervals, e.g. for gamma-ray bursts, can also be analysed. The likelihood is given by

\begin{equation}
\mathcal{L}(\boldsymbol{\theta}|D) = \prod_{p=1}^{N_{obs}} \frac{m_p^{d_p} \exp(-m_p)}{d_p!}\mrm{,}
\label{eq:likelihood}
\end{equation}

where $m_p$ is the model to describe (to be fitted to) the data $d_p$ per pointing $p$. The log-likelihood (C-stat) is then given by

\begin{equation}
\mathcal{C}(\boldsymbol{\theta}|D) = -2 \ln\mathcal{L}(\boldsymbol{\theta}|D) = -2 \sum_{p=1}^{N_{obs}} \left( d_p \ln m_p - m_p - \ln d_p! \right) \mrm{.}
\label{eq:loglikelihood}
\end{equation}

We will use the likelihood and related values (such as C-stat or $\chi^2$-values) and the degrees of freedom (dof: number of data points minus the number of fitted parameters) in a maximum likelihood fit as a goodness-of-fit criterion for our BG modelling method in Sec.~\ref{sec:fit_adequacy}. 

SPI data are typically dominated by instrumental BG\footnote{Only in a few cases during the INTEGRAL mission, transient sources appeared which out-shined the BG. One example would be the microquasar V404 Cygni \citep[e.g.][]{Rodriguez2015_v404,Roques2015_V404,Siegert2016_V404,Jourdain2017_V404} during revolutions 1554--1557.}, so that a subtraction in neither spectral, nor in the $\left\|d\right\| \times \left\|p\right\|$-dimension would provide reasonable results. Furthermore, there is no absolute BG model for SPI as there are variations in all dimensions which change according to the unpredictable solar activity. The goal is hence to obtain a description of the BG, from the data itself, but independent from the celestial source of interest. The key to such a BG model is found in the way, the data is modelled, and how the different model terms (can) influence each other:

\begin{equation}
m_p = \sum_{t} \sum_{j} R_{jp} \sum_{k=1}^{N_S} \theta_{k,t} M_{kj} + \sum_{t'} \sum_{k=N_S+1}^{N_S+N_B} \theta_{k,t'} B_{kp}\mrm{.}
\label{eq:spimodfit_model}
\end{equation}

Here, $M_{kj}$ is the k-th of $N_S$ sky images (celestial emission models) to which the instrumental response function (IRF, coded-mask shadowing), $R_{jp}$, is applied for each pointing $p$ and image element (e.g. pixel), $j$. The $N_B$ BG models $B_{kp}$ are independent of the IRF in each observation, which means the background is assumed completely independent from the shadowing of the mask and therefore independent of spacecraft repointings. Both model parts, sky and BG, can depend on time, and on different scales, $t$ for celestial sources, and $t'$ for the BG. While the BG timing depends on the instrument materials, being activated and decaying on various time-scales as a consequence of solar activity, the variability of the sky emission is only subject to the physics of the sources themselves. This means that the change of the BG amplitude, $\theta_{k,t'}$, depends on the process which resulted in the BG $\gamma$-ray photon. This may change on very short time scales, e.g. prompt emission after an intense dose of CR particles from a solar flare, or very long scales, e.g. from a radioactive build-up, when the creation of radioactive material happens on shorter time-scales than the decay time.

\subsection{Spectral background and response data base}\label{sec:SPI-BGRDB}

The SPI BGRDB by \citet{Diehl2018_BGRDB} spans the energy range between 20 and 2000~keV per INTEGRAL orbit, and consists of parameters for 383 $\gamma$-ray lines, each modelled with four parameters, and two continuum parameters, depending on the energy\footnote{This includes both data formats, the "single events" data (SE) from 20 to 1392~keV and from 1745 to 2000~keV, as well as the "pulse-shape-discriminated" data (PSD) from 490 to 2000~keV.}. In the energy range between 1 and 8~MeV\footnote{Above 2~MeV, the public available data are termed "high-energy events" (HE), for which the smallest energy binning is set to 1~keV, instead of 0.5~keV for the lower energies.}, the count statistics drops rapidly towards larger energies, so that longer integration times are required with increasing energy to obtain robust fitting results. For the BGRDB above 2~MeV, the time between two annealing periods is used to fit 614 $\gamma$-ray lines on top of a broken power-law-shaped continuum (Weinberger et al. 2019, in prep.). The low- and high-energy bands were subdivided into smaller energy regions, $e$, to limit the number of fitted parameters per fit.

As a function of energy, $E$, the SPI spectra, $F(E)$, per detector $d$, orbit $r$, and energy range $e$, are described by

\begin{eqnarray}
F_{der}(E) & = & C(E; \boldsymbol{c}_{der}) + \sum_{i} L_i(E;\boldsymbol{l}_{der}^i)\mrm{,~with} \\
\boldsymbol{c}_{der} & = & \left(C_{0,der},\alpha_{der}\right)\mrm{,~and}\nonumber\\
\boldsymbol{l}_{der}^i & = & \left(A_{0,der}^i,E_{0,der}^i,\sigma_{der}^i,\tau_{der}^i\right)\mrm{.}\nonumber
\label{eq:fit_fun_spectrum}
\end{eqnarray}

Here, $C(E;\boldsymbol{c}_{der})$ and $L(E;\boldsymbol{l}_{der}^i)$ describe the shape of the $\gamma$-ray continuum and lines with the parameter tuples $\boldsymbol{c}_{der} = (C_0,\alpha)_{der}$ and $\boldsymbol{l}_{der}^i = (A_0,E_0,\sigma,\tau)_{der}^i$, respectively, where $i$ is the index of the i-th $\gamma$-ray line in the band $e$. The parameters $C_0$ and $\alpha$ are the amplitude and the spectral index of the power-law function $C(E)$, normalised at the pivot energy $E_C$. This shape is motivated from the superposition of many power-law-like continuum processes in the satellite:

\begin{align}
C(E;C_0,\alpha) & = C_0 \left(\frac{E}{E_C}\right)^{\alpha} \label{eq:conti_full}
\end{align}

The parameters $A_0$, $E_0$, and $\sigma$ represent the amplitude, centroid, and width of a symmetric Gaussian function, with $\tau$ accounting for the detector degradation. The line shape that results from cosmic-ray bombardment is physically motivated \citep{Kretschmer2011_PhD}: As the regular lattice structure is deteriorated, the electronic structure changes, and charge carriers may be temporarily trapped. This leads to a time delay in the charge collection process, and thus in the charge pulse of the read-out electronics. The non-trapped portion of a charge cloud is proportional to $\exp(-\kappa x)$ \citep{Debertin1988_gamma}, where $x$ is the distance to Ge detector electrode, and $\kappa$ is the absorption coefficient. Since these parameters cannot be determined independently, a combined degradation parameter $\tau$ is used. The convolution of a symmetric Gaussian, $G(E;A_0,E_0,\sigma)$, with an exponential tail function, $T(E;\tau)$, leads to an asymmetric line shape, $L(E;A_0,E_0,\sigma,\tau)$ \citep[see also Eqs.~(3)--(6) in][]{Diehl2018_BGRDB}:

\begin{align}
G(E;A_0,E_0,\sigma) & = A_{0} \exp \left(- \frac{(E-E_{0})^2}{2\sigma} \right) \label{eq:gaussian} \\
T(E;\tau) & = \frac{1}{\tau} \exp \left( - \frac{E}{\tau} \right) \quad \forall E > 0 \label{eq:tail} \\
L(E;A_0,E_0,\sigma,\tau) & = (G \otimes T)(E) = \nonumber \\
& = \sqrt{\frac{\pi}{2}} \frac{A_{0} \sigma}{\tau} \exp \left( \frac{2 \tau (E-E_{0}) + \sigma^2}{2 \tau^2} \right) \nonumber \\
& \erfc \left( \frac{\tau (E-E_{0}) + \sigma^2}{\sqrt{2} \sigma \tau} \right) \label{eq:cls_function_full}
\end{align}

\citet{Diehl2018_BGRDB} finds that there are families of $\gamma$-ray lines in the instrumental background of SPI, which are characterised by their detector patterns. The authors define several groups of lines, of which we recap the most prominent: The "Ge-like" lines show higher count rates for inner detectors ($00$--$06$) with respect to outer detectors ($07$--$18$). These include excitation of nuclei in the SPI camera, such as Ge and Al, but also related and produced isotopes, such as Ga, Zn, and Mg. "Bi-like" lines show the opposite detector pattern, as their main source are in the materials of the SPI anticoincidence shield, made of bismuth-germanate (BGO). Related isotopes in the vicinity of Bi (e.g. Pb, Ra, Tl) and also isotopes from the actinide alpha-decay chains (e.g. Ac, Th, U) show this behaviour. To a lesser extent, also material from mountings (Ti, V), wires (Cu, Co, Fe), and other instruments aboard INTEGRAL (Cd, Te, Cs) show this pattern.

The most important finding from monitoring the long term trends in this spectral response data base is, that the patterns of individual isotopes stay constant on all time scales. This also means that isotopes which produces different line energies show the same pattern. Only detector failures change this behaviour - but then, the patterns are again constant. This is the prime assumption of how a self-consistent background model is constructed, and is further explained in Sec.~\ref{sec:general_approach}. Since the spectral shape changes with time (see above), the detectors may respond differently to this degradation, and the prompt and delayed background emission is different for different energies, the SPI BGRDB includes the information to reconstruct the expected spectral response of the SPI camera and predict the instrumental background as a whole at any point in time in high spectral resolution.

\section{Construction of the background model}\label{sec:bg_model_construction}

\subsection{General approach}\label{sec:general_approach}

At each specific energy, $E$, we model the BG by two components: photons from continuum processes, and from one (or more) $\gamma$-ray line-producing processes. In narrow energy bins, i.e. less than a few times the instrumental resolution\footnote{Between 20 and 8000~keV, the instrumental resolution is determined from narrow background lines, and ranges from 1.7 to 8.5~keV.}, we construct the BG by a combination of \textit{the continuum} and \textit{the lines}, i.e. combining all continuum-like processes and line-like processes to two individual BG models. Depending on the energy, either of the BG components may dominate, Sec.~\ref{sec:point_source_emission} (Fig.~\ref{fig:flux_npar_aic_crab}). Per each pointing, $p$, the factor $\sum_{j} R_{jp} M_{kp} \equiv S_{kp}$ changes for each sky component $k$, i.e. the detector pattern - which detectors are illuminated and how strong - changes with time / pointing. This is illustrated in Fig.~\ref{fig:sky_pattern} for the standard INTEGRAL $5 \times 5$ observation scheme of a point-like source. 

On the other hand, for a specific \textit{physical process} inside the satellite, the detector patterns from the BG, $B_{kp}$, do not change. For our two BG model components, $k=c$ for \textit{the continuum}, and $k=l$ for \textit{the lines}, hence follows: $B_{cp} = \mrm{const.} = B^c$ and $B_{lp} = \mrm{const.} = B^l$. What might change as a function pointing (=time), $p$, are the amplitudes $\theta_c(p)$ and $\theta_l(p)$. Therefore only these are determined in the maximum likelihood fit, following Eqs.~(\ref{eq:loglikelihood}) and (\ref{eq:spimodfit_model}), see also Secs.~\ref{sec:BG_changes} and \ref{sec:time_scale}.

Following these considerations \citep[see also][Sec.~2.4]{Diehl2018_BGRDB}, the SPI BG model for a particular detector, $d$, at a specific pointing, $p$, and energy bin, $e$, is written as

\begin{eqnarray}
B_{d,e,p} & = & \theta^c_{e,p} \times B^c_{d,e,p} + \theta^l_{e,p} \times \sum_{i=i(e)}^{N_{lines}(e)} B^l_{d,e,p} = \label{eq:line_one_bg_model} \\
& = & \theta^c_{e,p} \times C(\boldsymbol{c}_{d,e,p(r)}) \cdot t'^c_{e,p} + \nonumber\label{eq:line_two_bg_model}\\
& + & \theta^l_{e,p} \times \sum_{i=i(e)}^{N_{lines}(e)} L_i(\boldsymbol{l}_{d,e,p(r)}^i) \cdot t'^l_{i;e,p}\mrm{.}\nonumber
\label{eq:total_bg_model}
\end{eqnarray}

In Eq.~(\ref{eq:line_one_bg_model}), the BG is factorised into the fitted amplitudes for \textit{continuum} and \textit{lines}, $\theta^c_{e,p}$ and $\theta^l_{e,p}$, respectively, the spectral parts, and a temporal part. Here, individual pointings may be scaled by the same amplitude parameter (see Sec.~\ref{sec:time_scale}). For the \textit{lines} component, we sum over the individual lines $i$, which contribute to one energy bin, $e$. In principle, each line amplitude could be fitted individually, but this is only robust for strong BG lines. The spectral parameters per pointing, $\boldsymbol{c}_{d,e,p(r)}$ and $\boldsymbol{l}_{d,e,p(r)}^i$, depend on in which revolution the pointing occurred, hence $p(r)$\footnote{This is true because the SPI BGRDB was created on an orbit time scale. This can be generalised to arbitrary time bins in spectral parameter data bases, e.g. above 1~MeV where the parameters are being determined on "annealing" time scale, i.e. the average over half a year.}, and are fixed to the values from the SPI BGRDB in the maximum likelihood fit.

The detector pattern per energy and process is included as each detector is assigned a specific value according to the spectral response at that time and energy. Thus, $C_{d,e,p(r)}$ and $\sum_{i=i(e)}^{N_{lines}(e)} L_{i;d,e,p(r)}$ completely determine the spectral as well as detector ratio information required for BG modelling. The tracer functions $t'^c_{e,p}$ and $t'^l_{i;e,p}$ then allow for a relative weighting between pointings for \textit{continuum} and \textit{line} processes, respectively. These are parts of the specific BG modelling per science case, and thus are independent from the SPI BGRDB, hence do not follow $p(r)$. In general, each process can have its own variability, being prompt from cosmic-ray excitation and instantaneous de-excitation, or delayed when there is a longer lifetime of the produced isotopes included. The latter can lead to long-lasting radioactive decays, e.g. after a strong solar flare which created a lot of radioactive material (e.g. $\mrm{^{48}V}$ with a half-life time of 16 days), or to radioactive build-up when the decay-time is considerably larger than the production rate \citep[e.g. $\mrm{^{60}Co}$ with a half-life time of 5.27 years, cf.][]{Diehl2018_BGRDB}. The long-term trends are traced already by the SPI BGRDB as the amplitudes are determined on three-day time scales (Sec.~\ref{sec:BG_changes}), so that the remaining prompt BG emission processes at each energy can be traced by one temporal function, i.e. $t'^c_{e,p} = t'^l_{i;e,p} := t'_{e,p}$.

\subsection{Temporal behaviour}\label{sec:BG_changes}

At individual energies, i.e. for small energy bins - not integrating over the energy range of the physical process, the instrumental $\gamma$-ray line BG patterns, $B^l$, become a function of energy because of detector degradation. This does not mean that the physical processes change - the pattern per process is constant - but the pattern at a particular energy can change. This is caused by the degradation of individual detectors on different time scales due to their individual constitutions, behaviours, and reactions to particle irradiation. These changes in detector patterns versus energy and time are taken into account by using the SPI BGRDB. This data base traces the small but important changes, as it has been created on a three-day time scale for energies below 2~MeV \citep{Diehl2018_BGRDB}. For higher energies, i.e. lower BG rate, the BG count statistics is rarely sufficient to determine the gradual change in degradation, so that a mean value over half a year is determined.

The SPI BG amplitude of a certain instrumental process is in general unpredictable. In SPI data analysis, these BG variations are approximated at first order by "tracers", i.e. rates of onboard radiation monitors, such as the SPI anti-coincidence shield count rate, or the rate of saturating Ge detector events ($\gtrsim 15$~MeV; GeDSat). These rates trace the cosmic-ray particle flux that leads to instrumental $\gamma$-ray BG. The variations of these BG time series during one orbit are of the order of 1\%, i.e. $<1\%$ from pointing to pointing.

These approaches work well for all energy bins, i.e. as small as 0.5~keV or even 1~MeV continuum bands. But it is important to note that the described tracing relies on the prompt effect of cosmic-ray irradiation, which is true for many, but not all BG-generating processes in the satellite. Any BG process on longer scales is not represented by such a tracer, but instead is already implemented in the SPI BGRDB. In the special case of mid-term radioactivities, i.e. longer than several pointings, solar-flare-induced background lines or radioactive build-ups may be traced directly by the respective exponential decay law with the isotope's characteristic decay time \citep[cf. Fig. 15 of][showing the examples $\mrm{^{48}V}$ and $\mrm{^{60}Co}$]{Diehl2018_BGRDB}. Thus instead of relying on independent rates, the natural time scale of the process can also be used.

This tracing provides a good first-order description of the BG model variation on shorter time scales below one orbit and down to pointing-by-pointing. When the BG is fitted to the data per energy bin, however, the appropriate BG re-scaling has to be determined. This depends on energy, because the average BG count rate changes with energy due to different contributions, and strengths of \textit{continuum} and \textit{lines}, and the different intrinsic time-scales of the processes, be they prompt or delayed. This is discussed in Sec.~\ref{sec:time_scale}.

With the experience of hundreds of analysed X- and $\gamma$-ray sources during 16 years of the INTEGRAL mission, in different sky regions and $\gamma$-ray energies, it has become evident that for a large energy range between $\approx 200$ and $8000$~keV, the saturating Ge detector events (GeDSat) are sufficient to trace the inter-pointing variations. The 511~keV BG line, for example, follows strongly the rate of the side-shield assembly of the SPI anticoincidence shield \citep[SSATOTRATE][]{Skinner2014_511,Siegert2016_511}. Below 200~keV, the BG rate is very high and can often be determined on a pointing time-scale, so that no tracer is required at all if the source is also strong. Note that GeDSat does not necessarily explain the full variation from one pointing to the next, and requires re-scaling in our method. Likewise, at the steps of the SPI BGRDB, the BG must be carefully investigated and re-scaled (Sec.~\ref{sec:time_scale}).

\subsection{Discussion - why does this work?}\label{sec:why_does_this_work}

\begin{figure}[!ht]%
\includegraphics[width=\columnwidth]{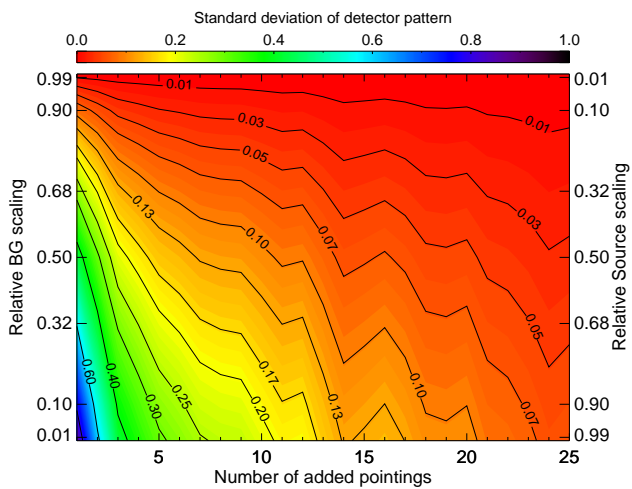}%
\caption{Variance of the detector patterns as shown for a typical $5 \times 5$-dithering pattern from Fig.~\ref{fig:sky_pattern}. For each added pointing, we calculate the standard deviation across the 19 Ge detectors, and also consider different source (right axis) to BG (left axis) ratios. For a specific ratio, the variance changes with the number of added pointings to the power of $-0.8$. Thus, even for strong source contributions, the mask pattern will smear out over the course of one INTEGRAL orbit.}%
\label{fig:pattern_smear_out}%
\end{figure}

In a single pointing, the mask patterns for sources in the field of view can be very strong, so that a determination of detector patterns from the BG ("BG response") is not possible on this time scale. Off-observations, for example at high latitudes, can be used if broad energy bins are analysed (continuum sources). For detailed spectroscopy in small energy bins, however, the BG patterns change smoothly with time due to detector degradation so that the actual spectra should be used to determine the BG. If many pointings of the same observation are combined, the source imprints due to the mask smear out, so that the resulting spectra per detector for a longer time scale appear as due to BG only. In Fig.~\ref{fig:pattern_smear_out}, we show how $S_{kp}$ smears out for a point source, observed with a $5 \times 5$-dithering strategy of INTEGRAL. For a source contribution of 10\% (90\%) to the total recorded spectrum, the cumulative pattern of BG plus source is varying by less than 1\% (10\%) after only 20 pointings. During one INTEGRAL orbit, 50--90 pointings are performed, so that the contribution of a residual mask pattern from even strong sources would be less than 1\%. This means that the SPI BGRDB \citep{Diehl2018_BGRDB} is mostly free of source contributions of any type, and that in this way, a BG model for single pointings can be constructed.

The detector patterns from instrumental BG continuum, $B^c$, are nearly flat around a value of 1.0 \citep{Diehl2018_BGRDB}. The limit, $\lim\limits_{N_p \rightarrow +\infty}{ \frac{1}{N_p} \sum_{p=1}^{N_p} S_{kp} }$, also evolves to a constant value of 1.0 across all detectors, so that any additional source contribution appears in the time-integrated spectra as an increased BG continuum. For individual pointings however, as analysed in a data set, thus, the amplitude of any source contribution can be determined in the fit because the expected patterns from BG and sky are now clearly different. The maximum likelihood fit including BG continuum, BG lines, and sources, will, as a consequence of this procedure, scale down \textit{the continuum} to "give way" to possible source counts.

As a result, the SPI BGRDB, fitted to the data, provides already a good first-order BG model for any INTEGRAL/SPI observation. For detailed and high-resolution spectroscopy of both continuum as well as $\gamma$-ray line sources - especially at fine energy binning - the characteristics of the parameter values in the data base must be investigated further to ensure consistency of the spectral description, in order to build a stable and robust BG model (see also Secs.~\ref{sec:BG_changes} and \ref{sec:test_cases}):

In the case of partial coding of sources during one INTEGRAL orbit, the relative counts of detectors at the rim of the SPI camera (parts of the outer detectors) may be increased on a per cent level (see above), so that the derived background patterns may be slightly skewed. This can also happen with strong sources in the fully coded field of view, leading to residual "average patterns" from these sources. The additional pattern, which is mixed with the true background pattern in the SPI BGRDB, is the sum of all detector responses from sources seen by SPI in a particular orbit (e.g. the sum of all patterns in Fig.~\ref{fig:sky_pattern}). This \textit{can} introduce weak coding in the partially coded field of view, but has marginal effects on the known sources. It affects mainly the energy range below $\approx 100~\mrm{keV}$, and can be accounted for by adding the expected "average pattern" as a third background model component.

\begin{table*}[!ht]%
\begin{tabular}{lrrrrccl}
Energy & Bin size & $N_{ebins}$ & $N_{obs}$ & $T_{exp}$ & Type                 & BGRDB     & Comments \\
\hline
508--514     &    6.0  &   1 & 12587 &  24.3  & diffuse/line         & o & inner Galaxy / also weak point sources\\
1795--1820   &    0.5  &  50 & 92867 & 201.0  & diffuse/line         & a & full sky \\
2500--3500   & 1000.0  &   1 & 11029 &  21.6  & diffuse/continuum    & a & inner Galaxy \\
\hline
50--100      &    2.0  &  25 &  3771 &  7.2  & point-like/continuum & o & \\
508--514     &    6.0  &   1 &  3771 &  7.2  & point-like/continuum & o & annihilation emission possible \\
1790--1840   &    0.5  & 100 &  3771 &  7.2  & point-like/continuum & o & no $\mrm{^{26}Al}$ expected\\
2500--3500   & 1000.0  &   1 &  3185 &  6.0  & point-like/continuum & a & \\ 
\hline
\end{tabular}
\caption{Summary of data sets used to show the performance of the BG modelling method. The first three columns define the analysed energy range and the chosen energy bin size in units of keV, and the resulting number of bins, $N_{ebins}$. The fourth and fifth column hold the number of pointings, $N_{obs}$, and the resulting exposure time, $T_{exp}$ in units of Ms for a working detector, from the chosen observations. In the "Type" column, the expected emission types, diffuse vs. point-like, and line vs. continuum emission, are listed. The "BGRDB" column shows on which data base was used to construct the background model, either per one INTEGRAL orbit (o), or per half a year (annealing, a). In the comments, additional characteristics of the data set or the expected emission are mentioned.}
\label{tab:data_sets}
\end{table*}

\section{Evaluation of the background model performance}\label{sec:fit_adequacy}

If the fluxes of the analysed sources are known, for example from previous analysis, or can be expected from theoretical considerations, it is possible to estimate the contribution of source counts in the data set. Consequently, the fit adequacy can be investigated, if the sources would be ignored. In general, the covariance between BG and source amplitude parameters is nearly zero, however will turn to negative values if the source contribution is weak or absent. This is shown and explained in Appendix~\ref{sec:covar_matrix}, Figs.~\ref{fig:correl_matrix_1795}--\ref{fig:corr_source_lines}, for the case of a line signal at 1809~keV (see also Sec.~\ref{sec:1809_bg_analysis}). As a consequence, the BG (time) scaling and the source contributions (also their time scaling if needed) should be optimised simultaneously.

As an absolute figure-of-merit to determine a fit's quality, we use the "modified $\chi^2_{\gamma}$-statistics" \citep{Mighell1999_chi2} as defined by

\begin{equation}
\chi^2_{\gamma} := \sum_{i=1}^{N_{obs}} \frac{\left[ d_i + min(d_i,1) - m_i \right]^2}{d_i + 1}\mrm{.}
\label{eq:mod_chi2}
\end{equation}

As an alternative to $\chi^2_{\gamma}$ and to distinguish between models, we use the Akaike Information Criterion \citep[AIC,][]{Akaike1974_AIC,Cavanaugh1997_AIC}. The AIC is defined as

\begin{equation}
AIC = 2 N_{par} - 2 \ln\mathcal{L}(D|\theta_i) + \frac{2N_{par}^2 + 2N_{par}}{N_{obs} - N_{par} -1}\mrm{.}
\label{eq:definition_AIC}
\end{equation}

This figure-of-merit is similar to the reduced $\chi^2$-value by penalising models with more parameters. Using relative likelihoods (absolute AIC differences, $\Delta\mrm{AIC}$), we identify on which time basis the BG has to be re-adjusted in the different cases of Sec.~\ref{sec:test_cases}. Note, that its absolute value has no proper meaning \citep{Burnham2004_AICBIC}. In general, the smaller the AIC, the more preferred a model is. It accounts for the appropriate likelihood of the data generating process, so that testing different assumptions on the BG (and source) scaling provides the most adequate number of BG (and source) parameters (Sec.~\ref{sec:time_scale}). The $\chi^2_{\gamma}$ and AIC values are \textit{calculated} from the best-fit $\ln\mathcal{L}$-value for reference. In general, one should avoid to rely on (reduced) $\chi^2$ estimations when dealing with Poisson-distributed data, however the $\chi^2_{\gamma}$-statistics is specifically adapted for this case \citep{Mighell1999_chi2}, and can provide a general measure.

\subsection{Data sets}\label{sec:data_sets} 

In order to illustrate different applications of the BG modelling method, we choose four energy regions to perform our analysis towards point-like and diffuse emission. We demonstrate the performance of our method on line-like emission at 511~keV, binned into one energy bin between 508 and 514~keV, as would be used for imaging. The second-strongest $\gamma$-ray line at 1809~keV of decaying $\mrm{^{26}Al}$ is analysed between 1795 and 1820~keV in 0.5~keV bins to also show a fine-binned but low-statistics case. Analysis of diffuse emission in a continuum band is presented between 2.5 and 3.5~MeV, summed into one bin of 1~MeV. For point-like emission, we also use a low-energy band, from 50 to 100~keV in 2~keV bins. We make use of Crab observations, performing tests in the same energy bands as for diffuse emission. A summary of the data sets and on which time basis the BGRDB is used, is found in Tab.~\ref{tab:data_sets}.

\subsection{Background changes with time}\label{sec:time_scale}

As described above, the relative weighting of background amplitudes between individual pointings can be determined by a tracer, but which may not fully cover the actual variations. In the following, we use pre-defined time steps to re-scale the BG model, as shown in Tab.~\ref{tab:time_scales}, and judge which scaling, i.e. how many fit parameters, are preferred. 

\begin{table}[!ht]%
\begin{tabular}{lrrrr}
Scaling        & Orbits             & Pointings                 & Days       & Hours \\
\hline
\verb|Const|   & $\infty$           & $\infty$                  & $\infty$   & $\infty$ \\
\multirow{4}{*}{\texttt{DetFail}}   & at 140 & at 5521          & at 1435.42 & - \\
               & at 214                      & at 10251         & at 1659.46 & - \\
							 & at 775                      & at 42443         & at 3337.50 & - \\
							 & at 930                      & at 52239         & at 3799.67 & - \\
\verb|Anneal|  & 60--70             & 1600--4500                & 170--210   & 4000--5000 \\
\verb|30n|     & 30                 & 1450--2150                & 90         & 2160 \\
\verb|20n|     & 20                 & 950--1450                 & 60         & 1440 \\
\verb|10n|     & 10                 & 400--800                  & 30         & 720 \\
\verb|5n|      & 5                  & 190--410                  & 15         & 360 \\
\verb|3n|      & 3                  & 110--250                  & 9          & 216 \\
\verb|2n|      & 2                  & 70--170                   & 6          & 144 \\
\verb|1n|      & 1                  & 50--90                    & 3          & 52--72 \\
\verb|24p|     & 1/3                & 24                        & 3/4--1     & 16--24 \\
\verb|12p|     & 1/6                & 12                        & 3/8--1/2   & 8--12 \\
\verb|6p|      & 1/12               & 6                         & 3/16--1/4  & 4--6 \\
\verb|3p|      & 1/24               & 3                         & 3/32--1/8  & 2--3 \\
\verb|2p|      & 1/36               & 2                         & 3/48--1/12 & 1--2 \\
\verb|1p|      & 1/72               & 1                         & 3/96--1/16 & 0.5--1 \\
\hline
\end{tabular}
\caption{Background re-scaling times used in the performance check. The number of orbits, pointings, days, and hours are presented for the largest used data set, the $\mrm{^{26}Al}$-case with 92867 pointings, comprising the full sky over 13.5 mission years. A \texttt{Const} background means, that only one background parameter per component is used for the entire data set, i.e. the background variability is fixed. The \texttt{1p} case allows for large variability as each observation pointing obtains its own background amplitude. The time nodes at which the four SPI detectors, 02, 17, 05, and 01 failed, are given in INTEGRAL Julian Days (IJD) and marked as \texttt{DetFail}. Intermediate background time scalings are chosen either inherent from the spectral background and response data base (\texttt{1n} and \texttt{Anneal}) or from selected times inbetween.}
\label{tab:time_scales}
\end{table}

In the case of diffuse emission, for instance, the expected detector pattern changes more smoothly than that for a point source. In addition, INTEGRAL/SPI data sets as they are typically analysed for diffuse emission cover very large areas in the sky, of which many are expected to include little or no source flux but only instrumental BG. For example, $\mrm{^{26}Al}$ is distributed dominantly in the plane of the Galaxy, but high-latitude observations are included in the data sets to get a better leverage on the absolute BG model. The data sets for large-scale diffuse emission thus include as many data/pointings as possible. In return this means, that the BG has to be re-scaled whenever emission from the sky - in certain regions, given by the emission models - is expected to change its contribution to the total by a significant amount. This could also be, when observing the same sky region at different mission times with different background levels. As a consequence, the sensitivity for diffuse emission with respect to analyses of point sources is reduced.

If a particular point source is investigated, the data sets are typically constrained to the region in which the source is located. When this same source is monitored from time to time, the BG level should be adjusted accordingly - not because the source flux might have changed\footnote{Of course, the source flux can also change, e.g. in the case of X-ray binaries. This just means that the fit is not stabilised by a constant source and varying BG any more, but rather independent for individual times.}, but because the BG level varies on longer time scales (e.g. solar cycle).

\section{Test cases}\label{sec:test_cases}

In the following Sec.~\ref{sec:diffuse_emission}, we will present the different diffuse emission cases and explain the choice of parameters. In Sec.~\ref{sec:point_source_emission}, we illustrate the findings for one point source, and highlight the differences as mentioned above.

\subsection{Diffuse emission - large data sets}\label{sec:diffuse_emission}

\subsubsection{Positron annihilation emission in the bulge}\label{sec:511_bg_analysis}

The 511~keV emission in the Milky Way from the annihilation of electrons with positrons is concentrated in the bulge region \citep[see][for a review]{Prantzos2011_511}. Here we choose only INTEGRAL observations in a central area around $(l/b)=(0^{\circ}/0^{\circ})$ with an extent of $\Delta l \times \Delta b = 15^{\circ} \times 15^{\circ}$. Due to the large field of view of SPI, also emission regions out to $|b|$ and $|l| \approx 20^{\circ}$ are included, and have to be modelled. We use the best-fitting emission model from \citet{Siegert2016_511} to characterise the 511~keV morphology in the bulge. This model consists of four 2D-Gaussian-shaped templates to represent the bulge and the superimposed disk. Here, we combine the individual components into one map.

\begin{figure}[!ht]
	\centering
				\subfloat[AIC vs. number of fitted BG parameters. \label{fig:npar_aic_511}]{\includegraphics[width=0.49\textwidth,trim=0.38in 0.65in 0.0in 0.49in,clip=true]{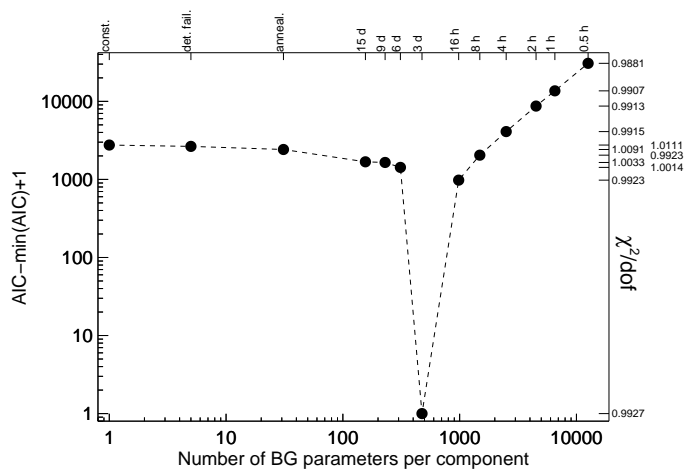}}\\
		  \subfloat[AIC vs. 511~keV flux. \label{fig:flux_aic_511}]{\includegraphics[width=0.49\textwidth,trim=0.38in 0.65in 0.0in 0.49in,clip=true]{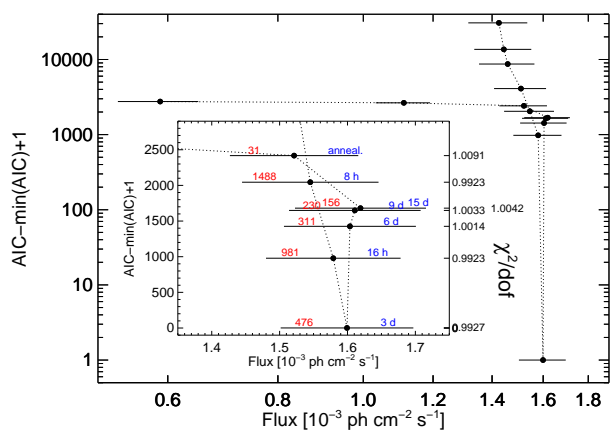}}
	  \cprotect\caption{Variation of the AIC in fits of the 511~keV emission band against the number of fitted background parameters per component (top). The bottom panel shows variations with the measured flux values for each fit, indicating the time scales in blue, and the corresponding number of fitted parameters in red. The corresponding reduced $\chi^2$-values are indicated on the right axis.}
\end{figure}

Clearly, a minimum is found at a BG time-scale of three days, i.e. one INTEGRAL orbit (Fig.~\ref{fig:npar_aic_511}). Between a time scale of 8~h and the annealing time scale (\verb|12p|--\verb|Anneal|), the measured flux shows only slight variations, indicating adequate and stable fits (Fig.~\ref{fig:flux_aic_511}). Very long (\verb|DetFail| / \verb|Const|) and very short ($\lesssim 8$~h; \verb|1p|--\verb|6p|) time scales, on the other hand, provide bad fits or over-fit the data, respectively, which leads to false flux values. For the optimum AIC at 3~d ($476 \times 2 = 952$ fitted BG parameters; \verb|1n|), the 511~keV flux is determined to $(1.6 \pm 0.1) \times 10^{-3}~\mrm{ph~cm^{-2}~s^{-1}}$, which is consistent with previous measurements considering a narrow and a broad bulge component, together with a disk in this region \citep[e.g.][]{Skinner2014_511,Siegert2016_511}.

The reduced $\chi^2$-value at the optimal AIC is 0.9927 for 203184 dof. This is $2.3\sigma$ from the canonically desired value of 1.0, and thus not "over-fitting" the data. In general in the 511~keV case, the fit, considering the $\chi^2$-statistics, is always acceptable: Between the \verb|Const| BG model with 2 parameters and a fit per each individual pointing with 25172 parameters, the reduced $\chi^2$ changes between 1.0111 and 0.9881, i.e. $+3.6\sigma$ and $-3.5\sigma$ from an optimal $\chi^2$-fit value of 1.0. The BG scaling closest to $\chi^2=1.0$ is found at \verb|2n|, i.e. one BG parameter every second INTEGRAL orbit (approximately six days). The minimum reduced-$\chi^2$ value is found for a fit per pointing, \verb|1p| (0.5--1.0~h), but which is largely penalised by the AIC, and thus represents an over-parametrised fit.

\subsubsection{Full-sky emission of radioactive $\mrm{^{26}Al}$}\label{sec:1809_bg_analysis}

The radioactive isotope $\mrm{^{26}Al}$ is produced in massive stars and ejected by winds and core-collapse supernovae. With a half-life time of 717~kyr, $\mrm{^{26}Al}$ traces the ongoing nucleosynthesis in the Milky Way. These nuclei decay via $\beta^+$-decay to an excited state of $\mrm{^{26}Mg}$, which is short-lived (476~fs) and de-excites to the stable ground state of $\mrm{^{26}Mg}$ by the emission of a $E_{lab} = 1808.63$~keV $\gamma$-ray photon \citep[cf.][for example]{Oberlack1996_26Al,Plueschke2001_26Al,Diehl2006_26Al,Diehl2010_ScoCen,Kretschmer2013_26Al,Bouchet2015_26Al,Siegert2017_PhD}. In this work, we use the Maximum Entropy Map from COMPTEL \citep{Plueschke2001_26Al} for our sky model to determine the BG scaling.

\begin{figure}[!ht]%
\includegraphics[width=\columnwidth,trim=0.32in 0.42in 0.59in 1.00in,clip=true]{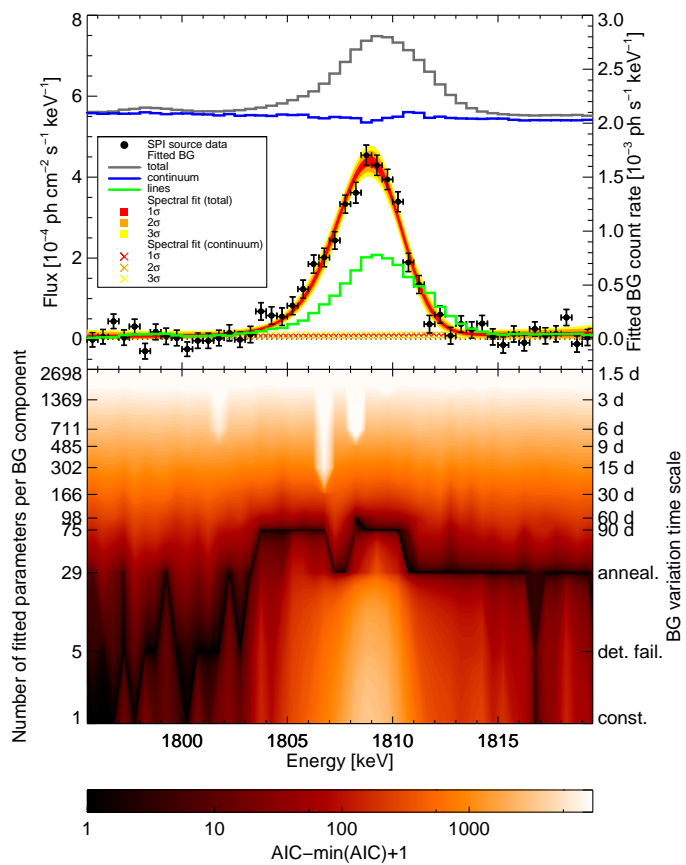}%
\caption{Study of background scaling in the $\mrm{^{26}Al}$ region. The bottom panel shows the AIC, similar to Fig.~\ref{fig:npar_aic_511}, but as a function of energy. From the optimal AIC at each energy bin, the flux is determined (black data points, upper panel, left axis). In addition, we show the fitted background (gray), divided into continuum (blue) and line (green) contributions (upper panel, right axis). To the black data points, a degraded Gaussian line on top of a power-law-shaped continuum is fitted, and shown as $1$, $2$, and $3\sigma$ uncertainty bands. The dotted line marks the zero flux level.}%
\label{fig:flux_npar_aic_1809}%
\end{figure}

Here, we use a 0.5~keV binning to resolve the 1809~keV $\gamma$-ray line between 1795 and 1820~keV. This allows us to show that the optimal BG scaling depends on energy, as the BG intensity rises when strong instrumental BG lines have to be taken into account. In Fig.~\ref{fig:flux_npar_aic_1809}, we show the result of our BG analysis in the $\mrm{^{26}Al}$ case for diffuse emission. While the energy bins containing the $\mrm{^{26}Al}$-line and a complex of strong BG lines (green), require a BG scaling between 30~d and half a year (\verb|30n|--\verb|Anneal|), the neighbouring continuum is sufficiently well fitted by only 1 to 29 parameters per BG component (\verb|Const|--\verb|Anneal|). Note, how also the weak instrumental BG line around 1797~keV is captured by our BG modelling method. In Appendix~\ref{sec:appendix_chi2}, we show the same analysis, based on the minimal reduced $\chi^2$-value in each energy bin. The major difference between the AIC and the $\chi^2$ profiles is in the erratic behaviour among bins, and the large spread in the choice of the optimal number of parameters. The line contribution also favours larger number of parameters - similar to the AIC analysis. The number of parameters in the $\mrm{^{26}Al}$ line is very large (5396), for which reason the errors bars on the derived fluxes are much larger in the optimal $\chi^2$-case than in the optimal AIC case. We compare the spectral parameters of the $\mrm{^{26}Al}$ region in Tab.~\ref{tab:26Al_spectral_fits}. All parameter are consistent, except for the line width, being as small as the instrumental resolution in the optimal $\chi^2$-case, and kinematically broadened for the optimal AIC case. Especially the 1.8~MeV line flux is consistent with previous studies of the full $\mrm{^{26}Al}$ sky \citep[e.g.][]{Bouchet2015_26Al,Siegert2017_PhD}.

\begin{table}[!ht]%
\begin{tabular}{lrrrrr}
Case              & $C_0$ & $\alpha$ & $I_L$ & $\Delta E_{peak}$ & $\Gamma_L$ \\
\hline
$\chi^2_{opt}$    & $3.9^{+3.6}_{-2.7}$ & $93^{+309}_{-119}$ & $1.57^{+0.14}_{-0.13}$ & $0.41^{+0.05}_{-0.05}$ & $3.17^{+0.14}_{-0.13}$ \\
$\mrm{AIC}_{opt}$ & $7.1^{+3.5}_{-3.6}$ & $35^{+92}_{-135}$  & $1.79^{+0.11}_{-0.10}$ & $0.27^{+0.05}_{-0.05}$ & $3.81^{+0.14}_{-0.14}$ \\
\hline
\end{tabular}
\caption{Spectral parameters of the $\mrm{^{26}Al}$-line region, selected for the optimal $\chi^2$-case and the optimal AIC case, $\chi^2_{opt}$ and $\mrm{AIC}_{opt}$, respectively. The units are $10^{-6}~\mrm{ph~cm^{-2}~s^{-1}~keV^{-1}}$, $1$, $10^{-3}~\mrm{ph~cm^{-2}~s^{-1}}$, $\mrm{keV}$, and $\mrm{keV}$ for the continuum amplitude $C_0$, the power-law index $\alpha$, the line flux $I_L$, the line centroid shift $\Delta E_{peak} = E_{peak} - E_{lab}$, and line width (FWHM) $\Gamma_L$, respectively.}
\label{tab:26Al_spectral_fits}
\end{table}

\subsubsection{Diffuse soft $\gamma$-ray continuum}\label{sec:3000_bg_analysis}

Above $\approx 100$~keV and up to several tens of MeV, the diffuse continuum emission in the Milky Way is dominated by inverse Compton scattering \citep{Strong2005_gammaconti}, with contributions from bremsstrahlung and positron annihilation  \citep{Beacom2006_511,Knoedlseder07,Strong2010_CR_luminosity,Lacki2014_SF_MeV_GeV}. Here, we use a continuum band between 2500 and 3500~keV to illustrate how our BG modelling method performs for diffuse continuum emission. We combine all photon events into one large energy bin of 1~MeV binwidth to increase the photon number statistics, and to illustrate the capabilities of our method in a broad band analysis.

\begin{figure}[!ht]%
\includegraphics[width=\columnwidth,trim=1.04in 0.69in 0.95in 0.97in,clip=true]{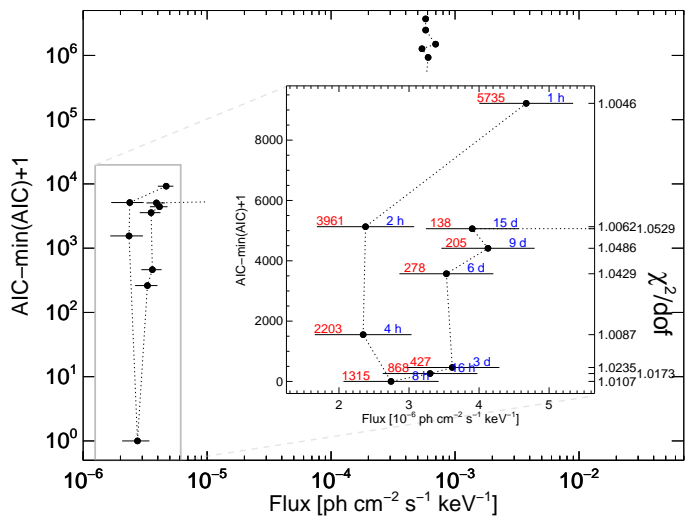}%
\caption{Variations of the measured flux between 2.5 and 3.5~MeV for each fit, similar to Fig.~\ref{fig:flux_aic_511}. The time scales are indicated in blue, and the corresponding number of fitted parameters in red. The corresponding reduced $\chi^2$-values are shown on the right axis of the inset.}%
\label{fig:flux_aic_3000}%
\end{figure}

For our analysis, we use the central part of the Milky Way, similar to Sec.~\ref{sec:511_bg_analysis}. Because the "high-energy" data sets of SPI, between 2 and 8~MeV, are very sensitive to solar activity, we removed periods in which solar flares strongly imprint and enhance BG intensities of nuclear de-excitation lines, such as $\mrm{^{2}H}$ at 2.2~MeV, $\mrm{^{12}C}$ at 4.4~MeV, or $\mrm{^{16}O}$ at 6.1~MeV. Consequently, the high-energy data sets are smaller than in the 511~keV case. We use an inverse Compton template from GALPROP \citep{Strong2011_GALPROP} to fit the morphology in the 2.5--3.5~MeV band.

The optimal AIC solution is found at a re-scaling time of $\approx 8$~h (\verb|12p|), corresponding to 2630 BG fit parameters for 176537 dof (Fig.~\ref{fig:flux_aic_3000}). The reduced $\chi^2$ value at this point is 1.0107, $3.2\sigma$ away from the desired $\chi^2/\nu=1.0$-solution. Based on the reduced $\chi^2$-value only, the optimal solution would be found at a 1~h BG scaling (\verb|1p|), but which is clearly disfavoured by the AIC, penalising too many fit parameters. The fitted flux values at $3.0\pm0.5$~MeV for acceptable fits range between $1.7$ and $5.3 \times 10^{-6}~\mrm{ph~cm^{-2}~s^{-1}~keV^{-1}}$. From these fits, a detection significance between $3.4$ and $7.0\sigma$ is found. At the optimum AIC, the diffuse continuum flux, based on the used inverse Compton template, is determined to $(2.7\pm0.7) \times 10^{-6}~\mrm{ph~cm^{-2}~s^{-1}~keV^{-1}}$. This is consistent with measurements from COMPTEL \citep{Strong1999_COMPTEL_MeV}, considering the different solid angles covered in the analyses.

\subsection{Point source emission - small and partitioned data sets}\label{sec:point_source_emission}

We use the Crab observations of INTEGRAL/SPI from 14 mission years. As a comprehensive example to be compared to the above cases, we show the energy regions from 50--100~keV in 2~keV bins, the positron annihilation line region between 508 and 514~keV in one 6~keV bin, the $\mrm{^{26}Al}$ region between 1790 and 1840~keV in 0.5~keV binning, and one 1~MeV bin between 2.5 and 3.5~MeV. Unlike in the diffuse emission examples, the Crab should only show continuum emission, and no $\gamma$-ray lines\footnote{Emission lines from decaying $\mrm{^{44}Ti}$ could be expected but are below the sensitivity limit for SPI.}. Similar to the high-energy case for diffuse emission, we apply special selection criteria to the Crab observations to avoid strong solar activity, for which reason also the Crab data set above 2~MeV is smaller than below. The choice of energy regions allows for direct comparison of the diffuse and line emission cases.

\begin{figure*}[!ht]%
\includegraphics[width=\textwidth,trim=0.60in 0.00in 0.10in 0.44in,clip=true]{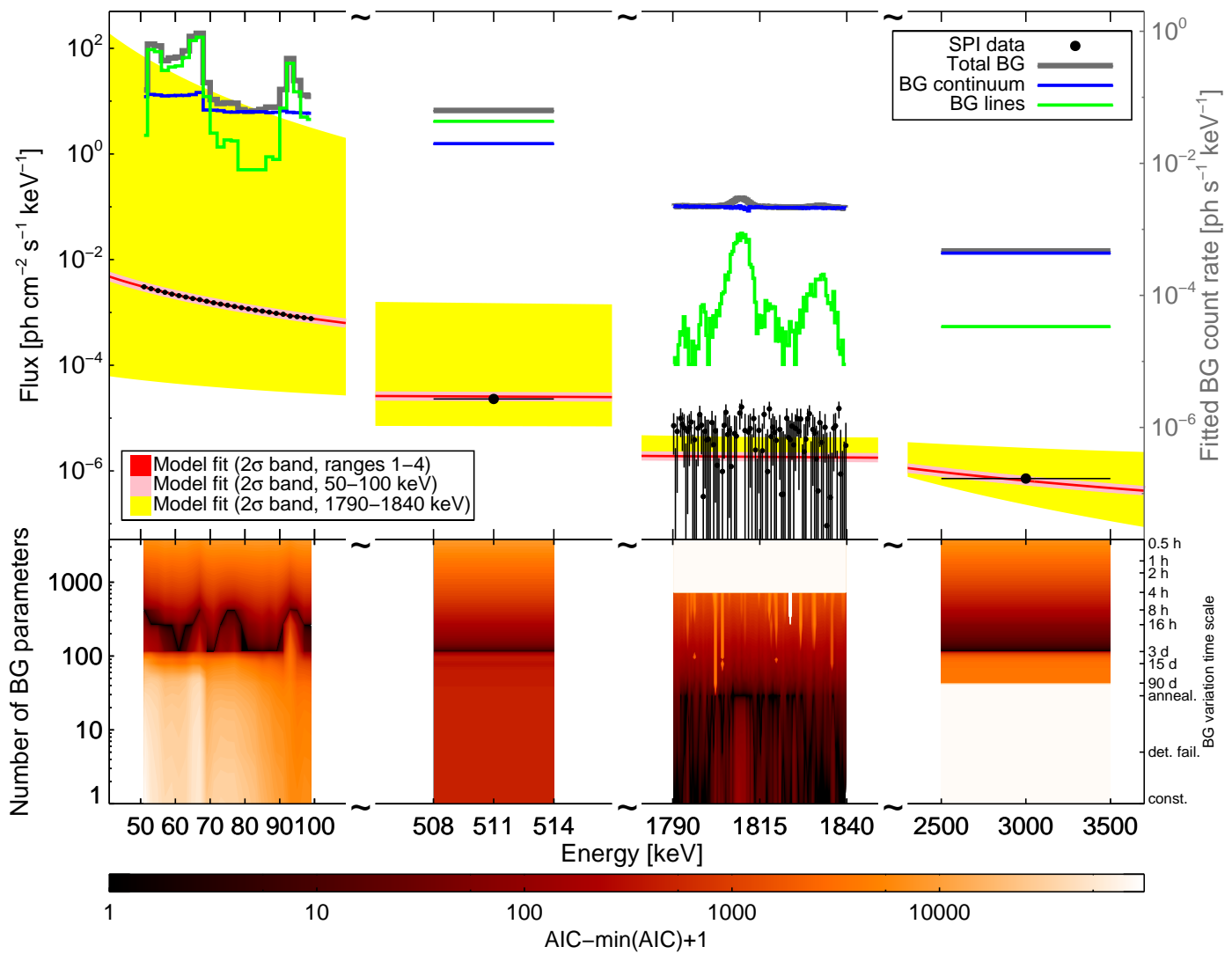}%
\caption{Study of background scaling in the Crab spectrum at different energies with varying energy binning, similar to Fig.~\ref{fig:flux_npar_aic_1809}. The bottom panel shows the AIC as a function of energy. From the optimal AIC at each energy bin, the flux is determined (black data points, upper panel, left axis). The fitted background is shown as gray histogram, divided into continuum (blue) and line (green) contributions (upper panel, right axis). The black data points are fitted by a power-law-shaped continuum, either only using the 50--100~keV range (pink band), only the 1790--1840~keV range (yellow band), or all four ranges together (red band).}%
\label{fig:flux_npar_aic_crab}%
\end{figure*}

In Fig.~\ref{fig:flux_npar_aic_crab},we show the results of our BG modelling method for the Crab and the time scaling analysis in four different bands with varying energy bin size. In the energy range, between 50 and 100~keV, it is evident that there is a large difference in BG re-scaling with time, as the BG flux changes rapidly with energy. Especially below $\approx 80$ and above $\approx 90$~keV, the instrumental BG \textit{line} contributions dominate over the instrumental BG \textit{continuum}, so that the BG scaling time is optimal between 4 and 8~h (\verb|6p|--\verb|12p|). In the \textit{continuum}-dominated region, 3~d (\verb|1n|) scaling and thus fewer parameters are enough for an optimal BG scaling. The resulting spectrum in this energy band is smooth and power-law like. Similar to the diffuse emission in the centre of the Galaxy, the BG amplitude in the 511~keV line band should be fitted every three days. There is no enhancement of a possible narrow 511~keV line in the spectrum, as determined from the spectral fit over all analysed energy bands (see below). Likewise, the energy band between 1790 and 1840, in which emission from $\mrm{^{26}Al}$ at 1809~keV could be expected, is sufficiently scaled once between each detector failure or annealing period (\verb|DetFail|--\verb|Anneal|) in the \textit{continuum}, and every 60--90~d (\verb|20n|--\verb|30n|) in the comparably strong instrumental \textit{line} complex between 1805 and 1813~keV. No $\mrm{^{26}Al}$ line is detected, which reinforces our method, being able to systematically suppress instrumental BG lines, and only reveals the source spectrum. At higher energies, between 2.5 and 3.5~MeV, the Crab flux is about the same order of magnitude as the diffuse emission in the galactic centre. However, the emission is concentrated into one point, so that the re-scaling must be less often than in the galactic diffuse case - here only once per three days (\verb|1n|). Also here, the difference in the uncertainties is evident between diffuse emission and point-like emission. While the galactic centre shows a flux uncertainty of $7.0 \times 10^{-7}~\mrm{ph~cm^{-2}~s^{-1}~keV^{-1}}$, the Crab emission is uncertain by $0.9 \times 10^{-7}~\mrm{ph~cm^{-2}~s^{-1}~keV^{-1}}$, even though the effective exposure times are about three times larger in the galactic centre (25~Ms) with respect to the Crab (7.5~Ms).

\begin{table}[!ht]%
\begin{tabular}{lrrr}
Energy range       & $A_0$                              & $\alpha$  & $E_m$~[keV]\\
\hline
50--100~keV    & $1.367(1) \times 10^{-3}$ & $-2.073(3)$        & $75$ \\
1790--1840~keV & $3.2(5) \times 10^{-6}$   & $-5^{+15}_{-22}$   & $1815$ \\
Full range     & $10.53^{+0.12}_{-0.17}$               & $-2.073(3)$        & $1$ \\
\hline
\end{tabular}
\caption{Spectral parameters of the Crab spectrum, derived from different energy bands. We normalise the spectra with $A_0$ in units of $\mrm{ph~cm^{-2}~s^{-1}~keV^{-1}}$ at different energies $E_m$ to avoid numerical problems.}
\label{tab:spec_params_crab}
\end{table}

We check for consistency in our analysis by fitting the Crab spectrum, using three different energy bands: (a) between 50 and 100~keV, (b) between 1790 and 1840~keV, or (c) all data points between 50 and 3500~keV. The fitting results are shown in Tab.~\ref{tab:spec_params_crab}, demonstrating consistency between the analyses, and further substantiating our BG modelling method. The Crab spectrum is determined to be $(10.53^{+0.12}_{-0.17}) \times E^{-2.073 \pm 0.003}~\mrm{ph~cm^{-2}~s^{-1}~keV^{-1}}$ between 50 and 3500~keV with a reduced $\chi^2$ value of 1.32 for 165 dof. This is consistent with previous SPI results reported for the Crab \citep[][with 1.2~Ms of observation time]{Jourdain2009_Crab}, and reduces the uncertainties by the increased amount of exposure.

\section{Conclusion}

In this work, we showed, how to construct BG models for INTEGRAL/SPI data from a data base of spectral parameters which characterise the response as well as the BG properties of individual components of the instrument. In particular we illustrate our method on a few characteristic examples, including point-like, diffuse, continuum, and $\gamma$-ray line emission.

Our method is based on long-term monitoring and understanding of instrumental BG, which is separated into \textit{continuum} BGs and $\gamma$-ray \textit{line} BGs. These physical components show individual, but characteristic detector patterns on the SPI detector array, and are constant over time. For individual energies (energy bins, not integrating over a $\gamma$-ray line, for example), the detector patterns change smoothly with time due to the change in their responses, mainly caused by detector degradation and the influence of solar activity. This is taken into account by our detailed spectral fits of INTEGRAL/SPI data per unit time \citep{Diehl2018_BGRDB}.

The time basis for these spectral fits must be chosen:
\begin{enumerate}
	\item long enough to smear out residual celestial patterns which are imprinted by the coded mask,
	\item short enough to trace gradual variations of detector responses, and
	\item long enough to accumulate enough statistics for reliable fits.
\end{enumerate}
Point 1 is fulfilled after 15--25 INTEGRAL pointings, depending on the strength of the source (Fig.~\ref{fig:pattern_smear_out}). We note that a correlation of our BG model with celestial emission is to be minimised by the way the BG is re-normalised in time, and want to point out that the more pointings are added for determining the BG response, the smaller the correlation will be. For example, a source contribution of 10\% (50\%) is smeared out to less than 1\% (5\%) after a standard $5 \times 5$-dithering pattern. This means one INTEGRAL orbit (50--90 pointings) would smear out the expected detector pattern of a single point source almost completely ($<3\%$ for any source strength; $<0.5\%$ for typical source contributions of less than 10\%), so that the correlation is minimal. At the same time scale, we find significant changes in the detector response ($\gamma$-ray line shape parameters) for the strongest $\gamma$-ray lines, so that point 2 is also fulfilled using one INTEGRAL orbit, or three days. Depending on the energy, i.e. BG count rate, the spectral fits are adequate (point 3) on a time scale of one orbit below energies of about 1--2~MeV. Above this energy, a spectral fit per detector requires more statistics to provide robust results. We choose the time span between SPI detector annealings to determine spectral parameters above 1--2~MeV. This reduces the accuracy for the line broadenings within half a year, but this can be accounted for by correcting for a linear trend using the orbit-by-orbit fits if required. We find, however, that fits per half-year period provide accurate results, as shown for the $\mrm{^{26}Al}$-case in (diffuse) line emission. If continuum sources are investigated and broader spectral bands are analysed, a spectral data base per orbit is also useful. 

We show that our BG modelling method is reproducible and also results in values for celestial sources which are in concordance with literature values. The optimal choice of BG re-scaling as a function of time, i.e. the number of parameters at each energy, depends on the emission type (diffuse or point-like), the spectral regime (low-energy or high-energy; dominated by instrumental \textit{continuum} or \textit{lines}), and the width of the energy bins analysed. We provide examples of how to determine adequate fits for INTEGRAL/SPI analysis, based on the Akaike Information Criterion. This penalises too many and disfavours too few parameters, and we compare the results to reduced $\chi^2$-values, which are commonly used in such analyses, but should be taken with care when using Poisson-distributed data with a low mean count rate. This BG method can be applied to a multitude of analysis cases, from short-term (e.g. gamma-ray bursts, state-variable X-ray binaries) to long-term (e.g. persistent sources, diffuse emission), continuum emission (e.g. synchrotron, bremsstrahlung, inverse Compton) and $\gamma$-ray line searches (e.g. decay, excitation, annihilation), as well as a combination of all of these at the same time. We show that our method eliminates most of the systematic uncertainties considering BG modelling with INTEGRAL/SPI.

\begin{acknowledgements}
This research was supported by the German DFG cluster of excellence 'Origin and Structure of the Universe'. The INTEGRAL/SPI project has been completed under the responsibility and leadership of CNES; we are grateful to ASI, CEA, CNES, DLR, ESA, INTA, NASA and OSTC for support of this ESA space science mission.
\end{acknowledgements}

\bibliographystyle{aa} 
\bibliography{alles} 

\appendix

\section{Covariance matrix of SPI maximum likelihood fits}\label{sec:covar_matrix}

\begin{figure}[!ht]%
\centering
\includegraphics[width=\columnwidth,trim=0.23in 1.65in 0.29in 1.57in,clip=true]{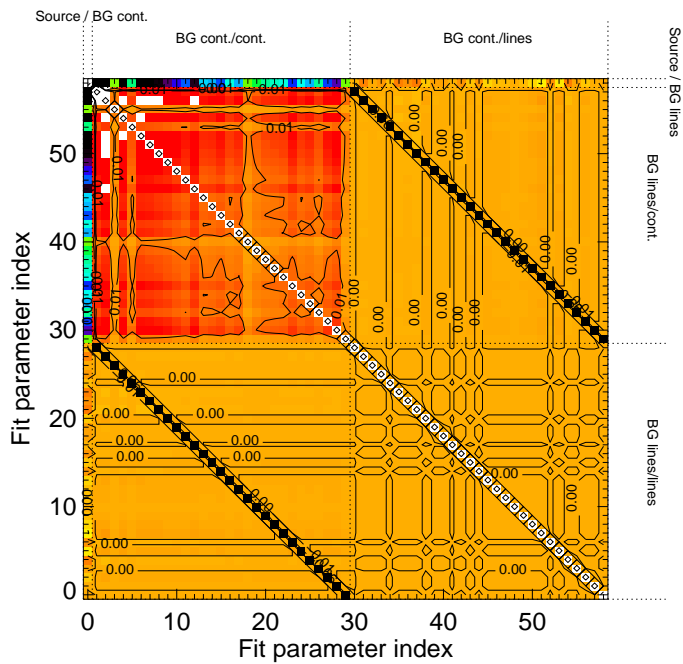}%
\caption{Correlation matrix of the three-component maximum likelihood fit with SPI in the energy bin 1795.0--1795.5~keV. The background component \textit{continuum} and \textit{lines} have been re-scaled 29 times each, corresponding to the time between two annealing periods (or detector failure and annealing period). The individual blocks are further explained in Figs.~\ref{fig:corr_conti_lines}--\ref{fig:corr_source_lines}.}%
\label{fig:correl_matrix_1795}%
\end{figure}

\begin{figure*}[!ht]
	\centering
			\subfloat[\textit{Continuum} vs. \textit{lines}. \label{fig:corr_conti_lines}]{\includegraphics[width=0.33\textwidth,trim=0.27in 1.97in 0.92in 0.92in,clip=true]{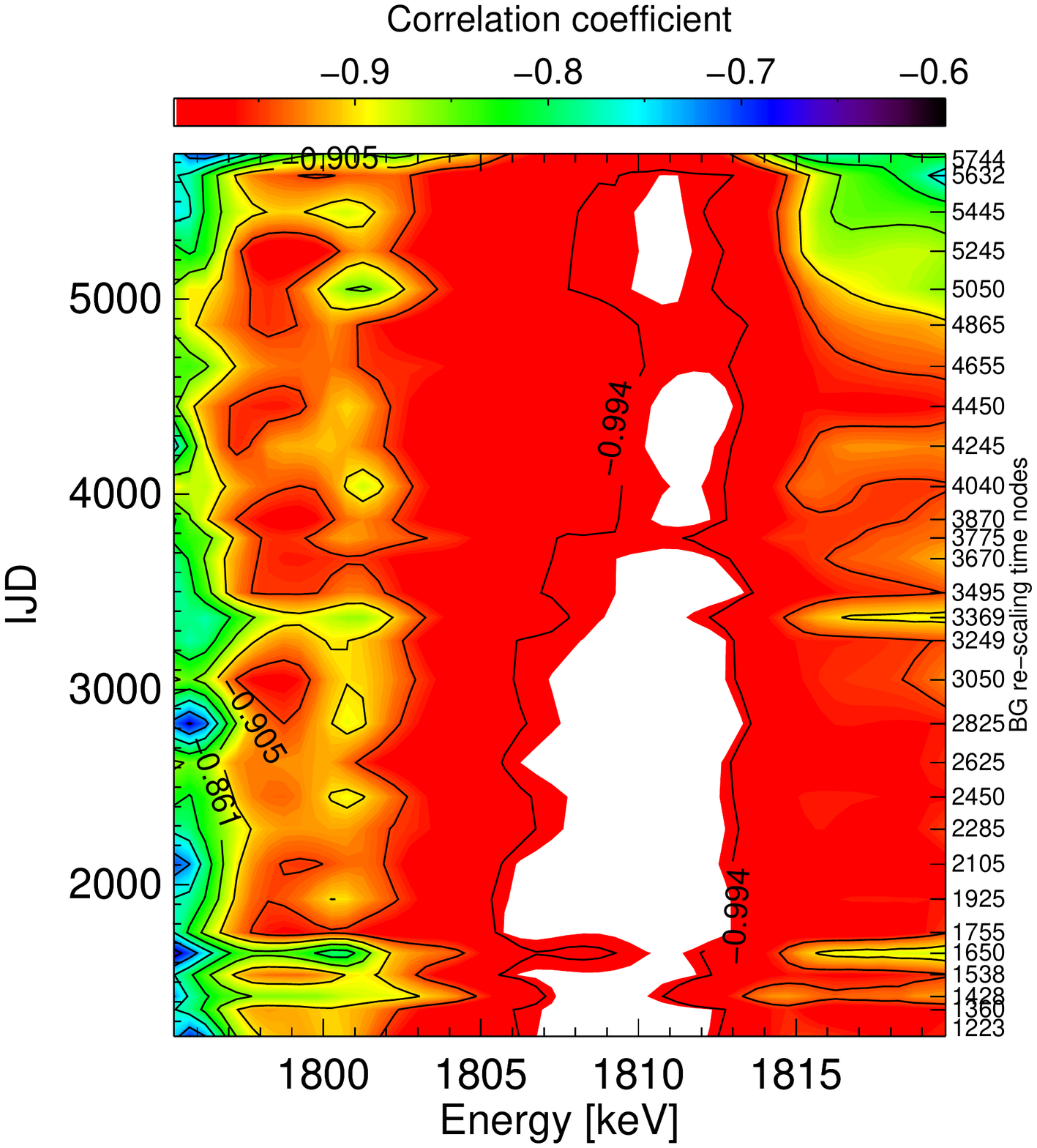}}~
		  \subfloat[Source vs. \textit{continuum}. \label{fig:corr_source_conti}]{\includegraphics[width=0.33\textwidth,trim=0.27in 1.97in 0.92in 0.92in,clip=true]{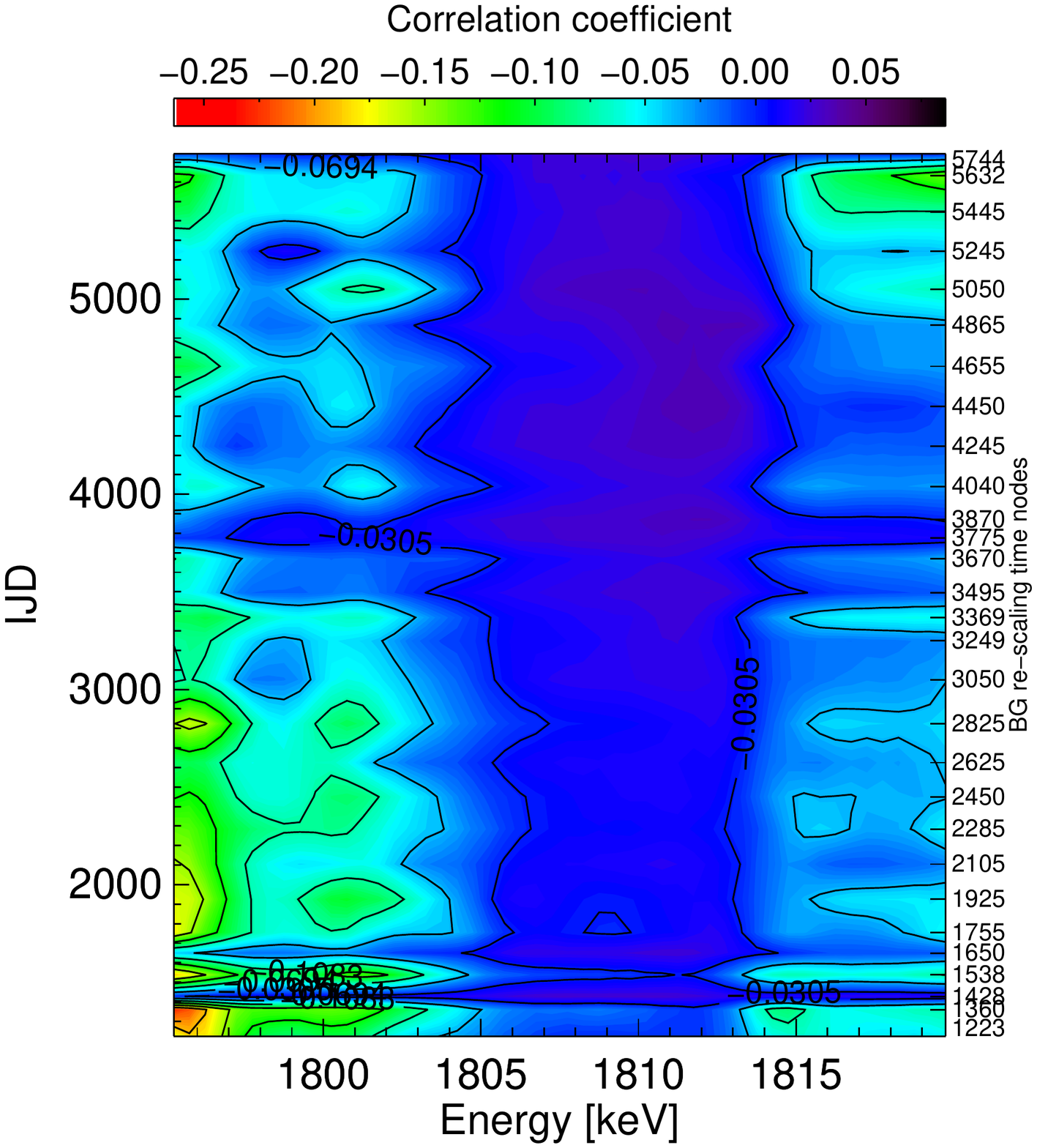}}~
			\subfloat[Source vs. \textit{lines}. \label{fig:corr_source_lines}]{\includegraphics[width=0.33\textwidth,trim=0.27in 1.97in 0.92in 0.92in,clip=true]{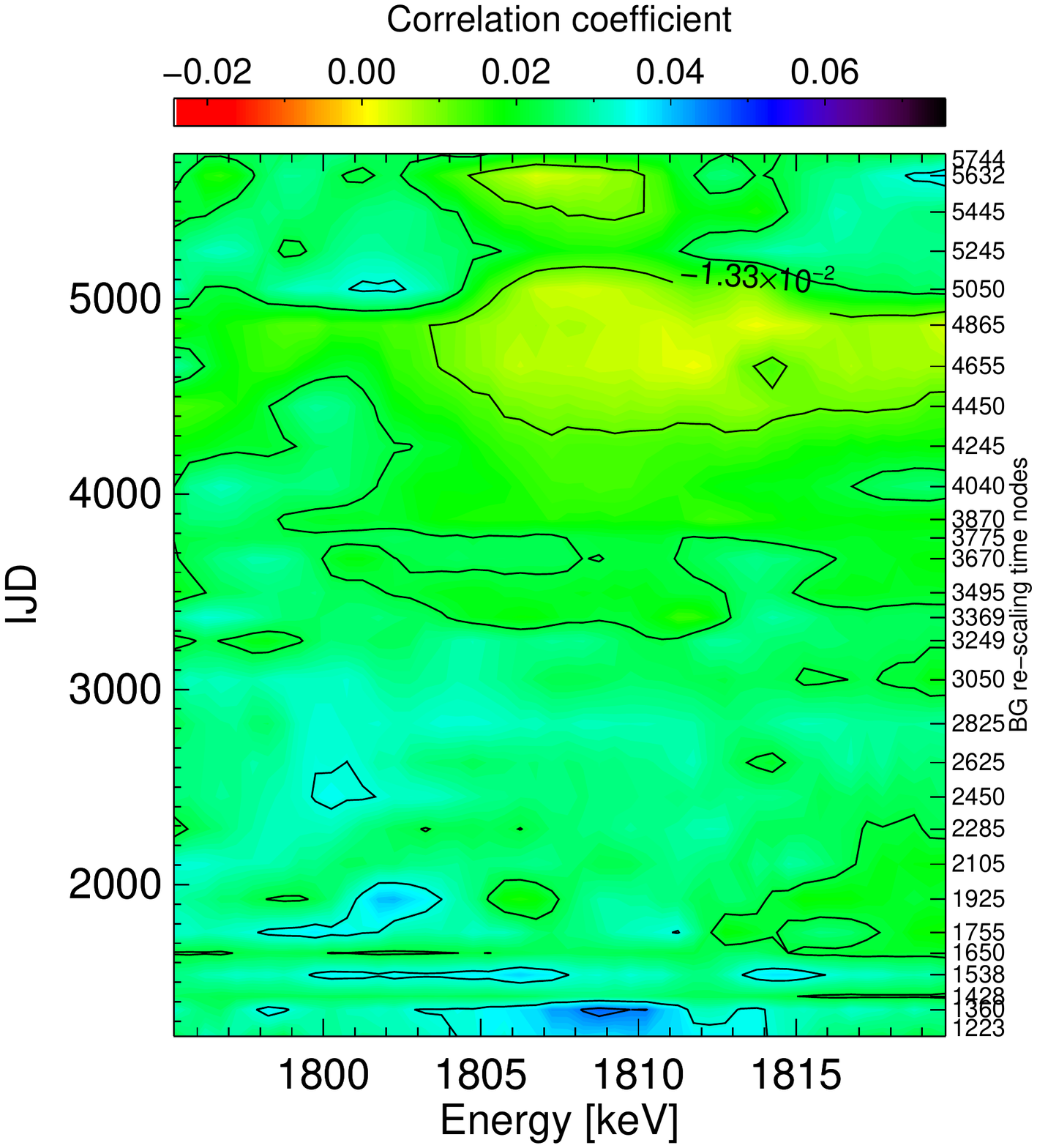}}
	  \cprotect\caption{Correlation matrix components as a function of energy and time for contemporaneous model components (no neighbouring fit parameters).}
\end{figure*}

As an example to illustrate the covariance matrix of a SPI maximum likelihood fit, we choose the case of $\mrm{^{26}Al}$ at 1809~keV (Sec.~\ref{sec:1809_bg_analysis}). Here, 29 BG fit parameters each for instrumental \textit{continuum} and \textit{lines} were used in addition to one fit parameter for the source. In Fig.~\ref{fig:correl_matrix_1795}, the covariance matrix of the fit in a single energy bin is shown. The correlations for consecutive time intervals (neighbouring fit parameters in one component) are in general around zero because these are only coupled by the one persistent source model in the fit. There are slightly higher correlations between neighbouring \textit{continuum} parameters than between \textit{continuum} and \textit{lines}, and \textit{lines} and \textit{lines}, because the BG detector patterns for the \textit{continuum} have less structure than \textit{line} patterns in general \citep[see also][]{Siegert2017_PhD,Diehl2018_BGRDB}. We show the correlations as a function of energy between source and \textit{continuum} (entries 1--29 in the top line in Fig.~\ref{fig:correl_matrix_1795}), source and \textit{lines} (entries 30--58 in the top line), and \textit{continuum} and \textit{lines} (diagonal in top right block of the matrix) in Figs.~\ref{fig:corr_conti_lines}--\ref{fig:corr_source_lines}. In each figure, the y-axis shows when the BG has been re-scaled in units of INTEGRAL Julian Days (IJD). The colour scheme shows the correlation coefficients as a function of energy and time.

It can be seen that the shorter times between detector failures and annealing periods (cf. Tab.~\ref{tab:time_scales}) lead to changes in the covariance structure for all components. The largest correlation between components is found between \textit{continuum} and \textit{line} BGs as expected from the data structure in such time intervals: between two time nodes, \textit{continuum} and \textit{line} BGs can be interchanged if the patterns are not different enough. Similar to a fit of a straight line, one component can replace the other to a certain extent, which leads to a strong anti-correlation. As described in Sec.~\ref{sec:general_approach}, the source shows the strongest correlations with the \textit{continuum} BGs because the detector patterns of smeared-out sources are closest to those of the BG continuum (Fig.~\ref{fig:corr_source_conti}). The correlations between line BGs and the source is flat and around zero because their share no similarities in detector space, even though they are centred at the same energies (Fig.~\ref{fig:corr_source_lines}).

\section{Background rescaling as a function of time using $\chi^2$}\label{sec:appendix_chi2}

\begin{figure}[!ht]%
\includegraphics[width=\columnwidth,trim=0.41in 0.39in 0.65in 0.97in,clip=true]{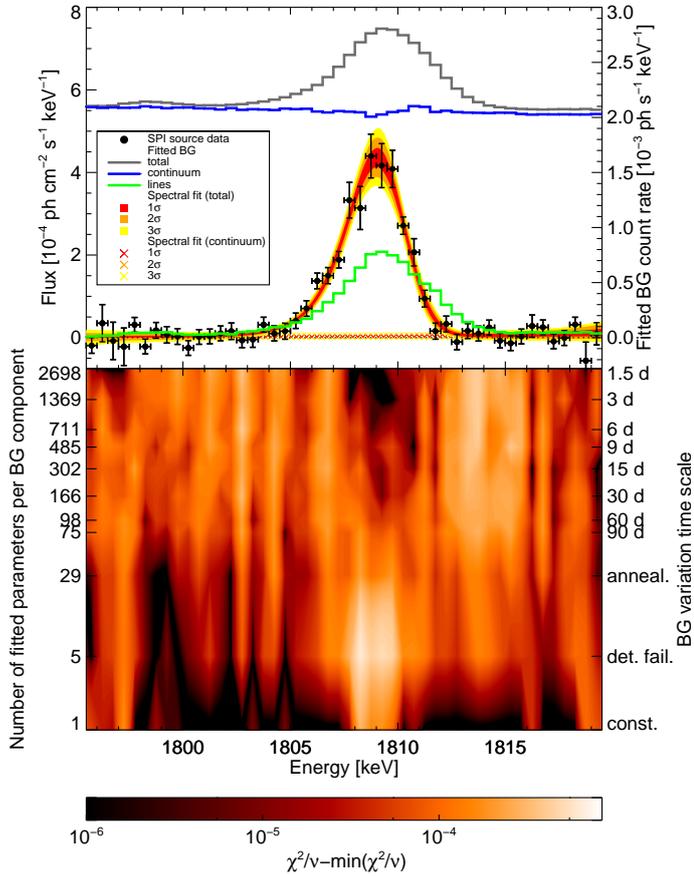}%
\caption{Study of background scaling in the $\mrm{^{26}Al}$ region as chosen from the optimal $\chi^2$ solution, similar to Fig.~\ref{fig:flux_npar_aic_1809}.}%
\label{fig:flux_npar_chi2_1809}%
\end{figure}

In general, $\chi^2$-distributions have expectation values equal to the number of degrees of freedom (dof), $\nu = N_{obs} - N_{par}$ \citep{Mighell2000_chigamma}. Their second moments (standard deviation) are $\sqrt{2\nu}$. Likewise, the reduced $\chi^2$-distribution, i.e. $\chi^2/\nu$, shows an expectation value of 1.0 with a standard deviation of $\sqrt{2/\nu}$. Note that these values are only correct for large Poisson mean values and over- or under-estimate the true value in the low-count regime. \citet{Andrae2010_chi2} discussed that the variance in $\chi^2$ means that even if the target value of $\chi^2/\nu = 1.0$ is not reached, the fit may still be considered "good". As the variance depends on the number of data points and the number of fitted parameters, allowed regions around the optimal value can be identified: for a data set with $N_{obs}$ data points and $\nu$ dof, a fit may be considered "good" if the reduced $\chi^2$ value falls in the range $\left[1-n\sqrt{2/\nu},1+n\sqrt{2/\nu}\right]$. Here, $n$ should be chosen according to the size of the data set, i.e. the number of data points $N_{obs}$. For large data sets, the chance of having individual $5\sigma$ outliers is larger than for small data sets, so that larger $n$ are adequate for larger data sets. This means that a reduced $\chi^2$ value of 1.2 can still be considered "good" for a data set with only 1000 dof, for instance, within 5$\sigma$, however extremely "bad" for a data set with $10^6$ dof, then being off by $\approx 140\sigma$. 

In Sec.~\ref{sec:time_scale}, we show the resulting fluxes for different data sets as derived from the optimal AIC solution. For our chosen examples with one broad energy bin each, Figs.~\ref{fig:flux_aic_511} and \ref{fig:flux_aic_3000} already contain the respective $\chi^2$-values for the different number of BG scaling parameters. It is worth to note that there is no trend for in- of decreased flux values when choosing the optimal AIC or optimal $\chi^2$ solution: In the 511~keV case, the fluxes are $F_{\chi^2}^{511} = (1.43\pm0.11) \times 10^{-3}~\mrm{ph~cm^{-2}~s^{-1}}$ and $F_{AIC}^{511} = (1.60\pm0.10) \times 10^{-3}~\mrm{ph~cm^{-2}~s^{-1}}$ for 25172 and 952 BG fit parameters, respectively. For the broad bin between 2.5 and 3.5~MeV, the solutions yield $F_{\chi^2}^{3000} = (4.7\pm0.7) \times 10^{-6}~\mrm{ph~cm^{-2}~s^{-1}~keV^{-1}}$ and $F_{AIC}^{3000} = (2.7\pm0.7) \times 10^{-6}~\mrm{ph~cm^{-2}~s^{-1}~keV^{-1}}$ for 11470 and 2630 parameters, respectively.

\end{document}